\documentclass[a4paper,fleqn,usenatbib]{mnras}
\usepackage[T1]{fontenc}
\usepackage{ae,aecompl}
\usepackage{multirow,ulem}

\usepackage{graphicx}	% Including figure files
\usepackage{amsmath}	% Advanced maths commands
\usepackage{amssymb}	% Extra maths symbols
\usepackage{natbib}
\usepackage{breqn}
\usepackage{amsmath}
\usepackage{siunitx}
\title[Dust attenuation in galaxies: main regulators]{Star-dust geometry main determinant of dust attenuation in galaxies}
\author[Sachdeva and Nath]
{
Sonali Sachdeva$^{1}$, Biman B. Nath$^{1}$ \\
\footnotesize \it $^{1}$Raman Research Institute, Sadashivnagar, Bengaluru 560080, India; \href{mailto:sonali@rri.res.in}{sonali@rri.res.in}\\ 
}
\date{Accepted XXX. Received YYY; in original form ZZZ}

\pubyear{2020}

\begin{document}
\label{firstpage}
\pagerange{\pageref{firstpage}--\pageref{lastpage}}
\maketitle

%%%%%%%%%%%%%%%%%%%%%%%%%%%%%%%%%%%%%%%%%%%%%%%%%%%%%%%%%%%%%%%%%%%%%%%%%

\newcommand{\3}{\ss}
\newcommand{\n}{\noindent}
\newcommand{\eps}{\varepsilon}
\def\be{\begin{equation}}
\def\ee{\end{equation}}
\def\ba{\begin{eqnarray}}
\def\ea{\end{eqnarray}}
\def\de{\partial}
\def\msun{M_\odot}
\def\div{\nabla\cdot}
\def\grad{\nabla}
\def\rot{\nabla\times}
\def\ltsima{$\; \buildrel < \over \sim \;$}
\def\simlt{\lower.5ex\hbox{\ltsima}}
\def\gtsima{$\; \buildrel > \over \sim \;$}
\def\simgt{\lower.5ex\hbox{\gtsima}}
\def\etal{{et al.\ }}

%%%%%%%%%%%%%%%%%%%%%%%%%%%%%%%%%%%%%%%%%%%%%%%%%%%%%%%%%%%%%%%%%%%%%%%%%
\begin{abstract}
Analysing a large representative sample of local galaxies (8707), we 
find that the variation in the shape of their dust attenuation curves is driven 
primarily by their structure, i.e., distribution of stars (and dust) within them. 
The attenuation curve for spheroid dominated galaxies, as compared to the disc 
dominated ones, is nearly twice as steep. Both structural types cover distinct 
ranges of attenuation slope values. Similar findings are reflected in the case of 
star-forming and passive galaxies. Spheroids and passive galaxies witness minimal 
attenuation in the optical compared to UV wavelengths underlining the lack of dusty 
birth-clouds that define complex star-dust geometry. The distinction in the 
attenuation properties of spheroids and discs is maintained in each stellar mass 
range emphasising that structure is the primal cause of variation. However, within a 
structural group, the attenuation curve becomes shallower with both the increase in 
total stellar mass and optical depth of the galaxy. Overall, with the extinction 
curve fixed to be the same for all galaxies, the star-dust geometry emerges to be 
the prime determinant of the variation in their attenuation properties.
\end{abstract}


\section{Introduction}

Dust, produced according to the star formation history of a galaxy, is a vital player 
in its ongoing and future star formation activity \citep{Lutz14,MadauandDickinson14}. 
Unravelling the connection between this dust and other physical properties of the 
galaxy is, therefore, a powerful diagnostic for the processes involved in its evolution \citep{Dunneetal11,Santinietal14}. The properties of dust inside a galaxy are deduced 
by analysing its affect on the galaxy's stellar emission over the spectrum 
\citep{Calzetti97,Draine03}. This affect is encapsulated by the dust attenuation curve 
of the galaxy that gives its optical depth as a function of wavelength. Constant 
efforts are made to compute these curves for galaxies in an increasingly accurate 
manner as it is a prerequisite for correct estimation of galaxies' stellar properties 
\citep{Walcheretal11,Conroy13}.

Several methods have been applied, over the years, for the estimation 
of these curves \citep[see review by][]{SalimandNarayanan20}. The method that has 
lately gained traction for providing most reliable attenuation curves for galaxies 
individually is the `model method'. This involves the creation of a library of 
dust-free SEDs spanning a large range of star formation histories and metallicities 
through the usage of stellar population synthesis models. The dust-free SEDs are 
then attenuated and compared with observed SEDs to obtain estimates of galaxy 
parameters and dust content. The main critique of this method has been the usage of 
a single dust prescription for computational ease and to break the formidable 
dust-metallicity-age degeneracy. In recent years, modellers have successfully 
resolved this issue by amending the procedure in such a way that best-fit dust 
attenuation parameters are also extracted along with other galaxy properties 
\citep{Lejaetal17,Salimetal18}. The amendment includes improvisation in the 
Bayesian MCMC fitting technique \citep{Boquienetal19,Johnsonetal19} and inclusion 
of IR observations to constrain total dust emission. \citet{Salimetal18} 
implemented this `energy balanced SED fitting' technique on a quarter million 
local galaxies with far-UV to far-IR observations. They have been the first ones 
to obtain a detailed set of dust attenuation parameters for galaxies individually, 
covering a wide stellar mass and star formation activity range, i.e., including 
passive sources as well.

This is an important procurement, not only for the accurate estimation of stellar 
properties but also to decipher the forces regulating the attenuation curves in 
galaxies. For a thorough probe, the attenuation parameters of a galaxy need to be 
compared with all its other defining properties, for a statistically large 
representative sample \citep{SalimandNarayanan20}. Although several studies have 
attempted this \citep{Wildetal11,Battistietal16}, they were constrained by either the 
choice of their sample or the availability of the whole set of attenuation parameters 
and other physical properties for each galaxy individually.

This constraint is lately getting resolved with the growing number of surveys focused 
on small extra-galactic patches, along with the attempts to measure all defining 
properties of galaxies in those patches. For example, for Stripe 82 - a 270 square 
degrees area in the sky - spectroscopic redshifts, deep images and spectra are 
available for tens of thousands of local ($z_{spec}<$0.3) galaxies 
\citep{Dawsonetal13,Annisetal14}. Utilizing that, \citet{Bottrelletal19} has computed 
the detailed set of structural parameters for all galaxies ($\sim$17000) in this area, 
in optical wavebands. Structural parameters in the infrared ($K_S$ band), reflecting 
the underlying mass distribution, are also available for a subset of the full sample 
from our earlier work \citep{Sachdevaetal20}. In addition to that, all 
defining dust attenuation and stellar parameters are available for these galaxies 
from GSWLC2 \citep[GALEX-SDSS-WISE Legacy Catalog 2,][]{Salimetal18}. Thus, detailed 
structural, stellar and attenuation curve defining parameters - understood to be the 
most accurate till date - are available for all ($\sim$17000) galaxies in the Stripe 
82 region. In this paper, we employ this data to perform an in-depth investigation of 
the processes and properties responsible for the observed variation in dust 
attenuation parameters of galaxies covering a wide stellar mass 
($7.5<\log M_*<12.5 M_{\odot}$) and star formation activity 
($-4<\log SFR<2 M_{\odot}/yr$) range.

\section{Data}

Stripe 82 is a 270 degree square strip along the celestial equator with $\sim2$ mag 
deeper SDSS imaging in all optical bands ($ugriz$) compared to the SDSS Legacy survey 
data \citep{Jiangetal14}. In addition to deep optical imaging, it has spectroscopic 
redshift measurements from BOSS \citep{Dawsonetal13}, deep UV observations from GALEX 
\citep{Bianchietal17} and near to far IR coverage from Spitzer \citep{Papovichetal16}, 
Herschel \citep{Vieroetal14} and VISTA \citep[VICS82,][]{Geachetal17} amongst others. 

Recently, \citet{Bottrelletal19} computed the whole range of multi-component 
structural parameters for all local (16908, $z<$0.3, $m_r<$17.77) galaxies in this 
field using deep co-added SDSS images. They have fitted both a single component 
(S\'ersic profile) model and a two component (S\'ersic profile $+$ exponential disc) 
model on each galaxy using a Bayesian maximum-likelihood optimisation 
algorithm \citep[GIM2D,][]{Simard98}. They demonstrate that fitting of galaxies 
using deeper images results in more accurate measurements mainly due to the 
reduction in systematics associated with background noise.

The dust attenuation parameters for this sample are available from GSWLC2 
\citep{Salimetal18}. Measurements made in GSWLC2 are based on the 
`model method' which involves the adoption of a parametric form for the attenuation 
curves. To avoid any potential bias stemming from this adoption, the slope of the 
assumed curve \citep{Calzettietal00} is allowed to deviate in the fitting process. 
Based on the relationship between stellar and nebular reddening, different 
attenuation levels are assumed for young and old populations, which is argued to be 
a more physically relevant approach as young stars are embedded in birth-clouds 
\citep{Salimetal18}. However, it is noted that this assumption may 
affect the steepness of the derived curves. An important improvisation over earlier 
efforts is the inclusion of IR luminosity (from WISE and HERSCHEL) that acts as a 
constraint on the total dust emission, thus allowing the attenuation curves to be 
fitted freely by the advanced software \citep[CIGALE,][]{Salimetal18,Johnsonetal19}. 
The stellar parameters of galaxies are estimated with regard to the best fit 
attenuation curves.

Thus, both attenuation and structural parameters are available for our chosen sample 
of local Stripe 82 galaxies (16908), covering a wide stellar mass and 
star formation activity range. For this work, we select only those galaxies that 
have been observed in both UV and IR bands, i.e., have $flag_{uv}>0$ and 
$flag_{midir}>0$ in the GSWLC2 catalog. This selection ensures that our sample 
(8707 galaxies) has the most accurate attenuation measurements. Note that the range 
and representative quality of the full sample is not affected by this selection. In 
our previous work \citep{Sachdevaetal20}, we focused on a subset of the total sample 
(1263 galaxies) to extract multi-component structural parameters in the near-IR 
range using $K_S$ band images from the VICS82 survey \citep{Geachetal17}. This subset 
allows us to examine if the structure related findings apply to the distribution of 
stellar mass or only the bright optical light within a galaxy. Thus, we have an ideal 
sample to analyse the cause of variation observed in the attenuation curves of galaxies.

\begin{figure*}
	\includegraphics[width=58 mm]{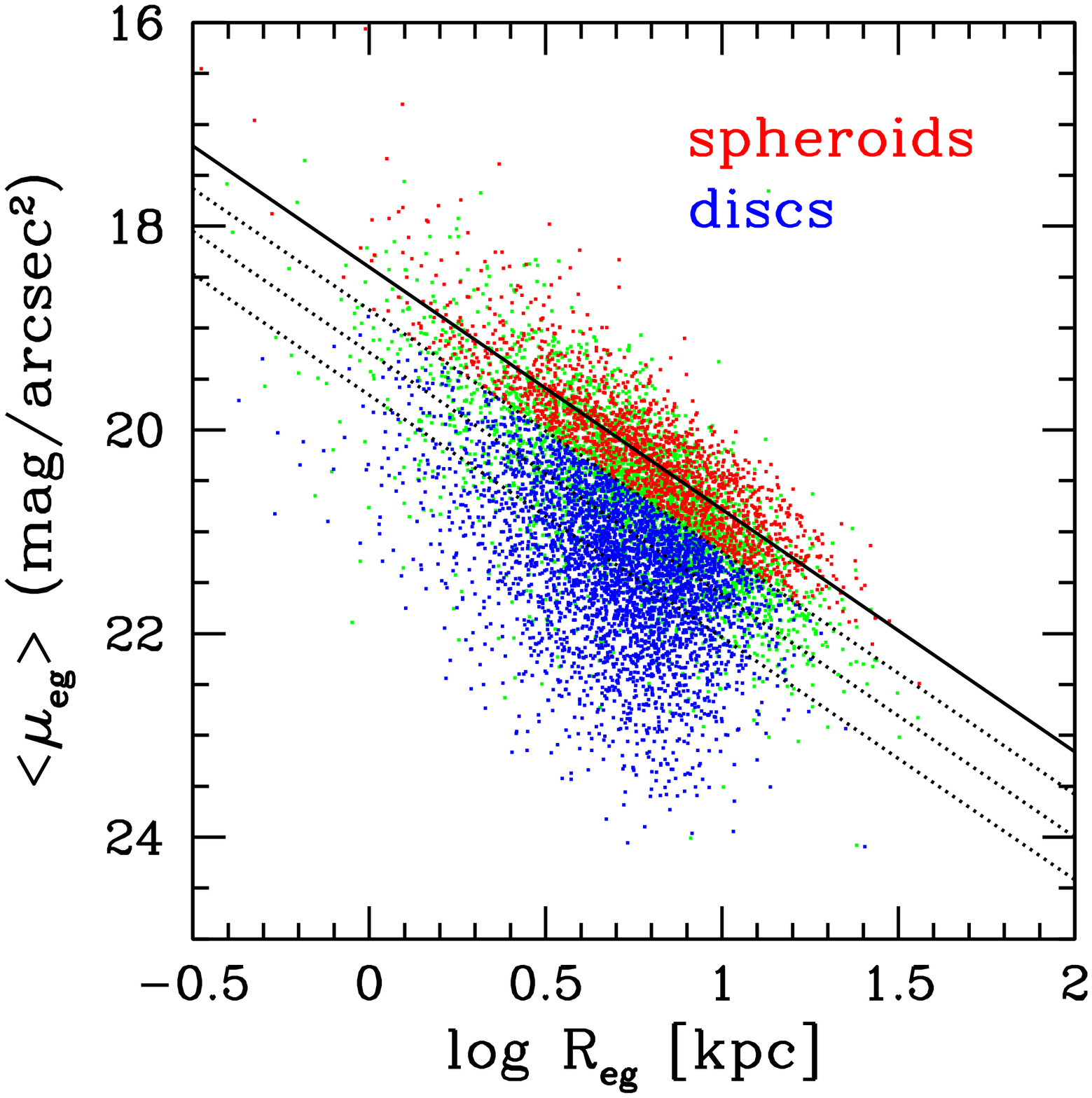}
	\includegraphics[width=58 mm]{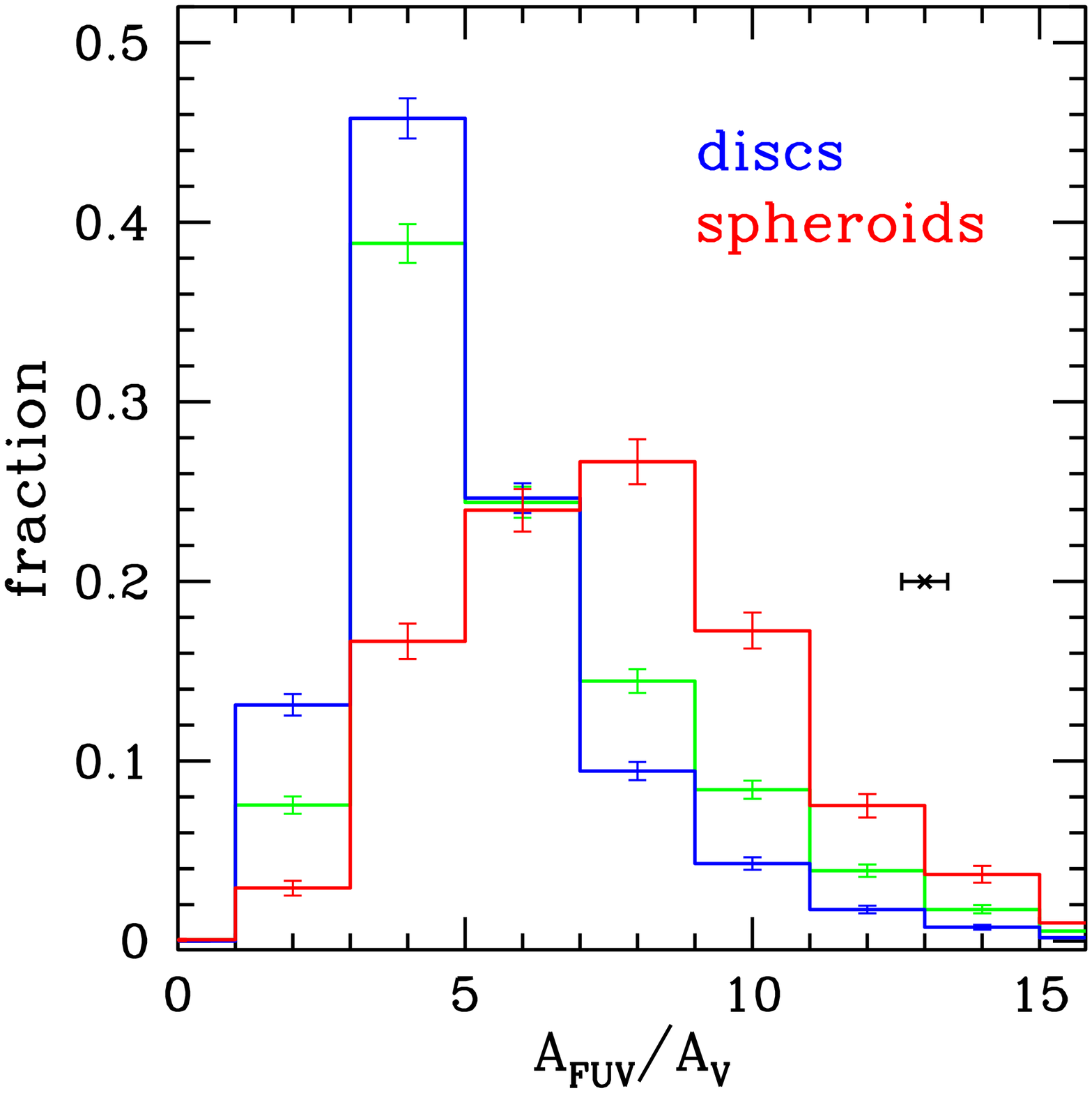}
	\includegraphics[width=58 mm]{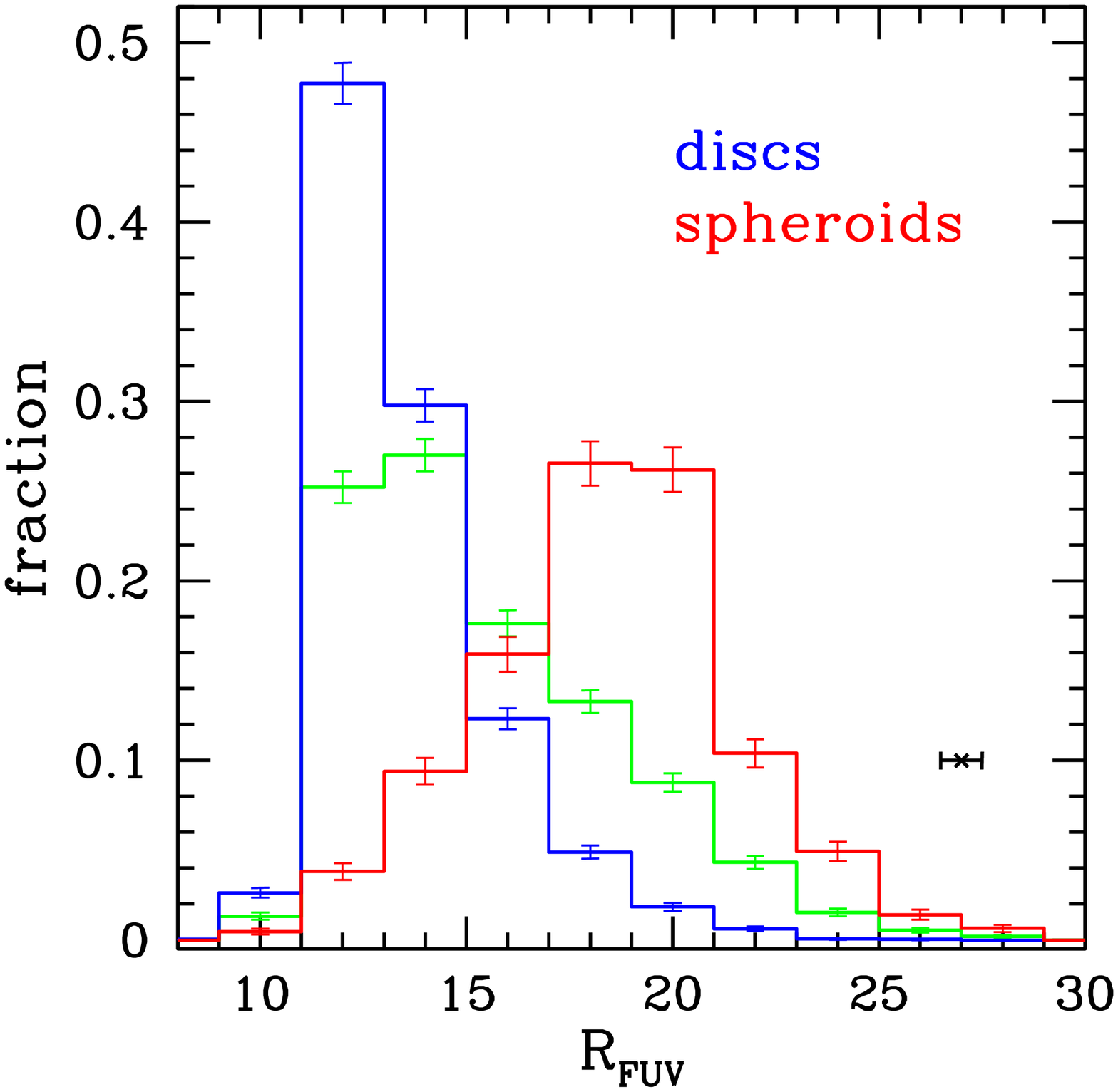}
    \caption{{\bf Structural classification} of galaxies is shown in the left-most 
    plot. The Kormendy Relation known for local elliptical galaxies 
    \citep{Longhettietal07,Sachdevaetal19} and its lower $\sigma$ boundaries are 
    marked with solid and dashed lines. Galaxies with global S\'ersic index ($n_g$) 
    more than 3.5 and placement within 1$\sigma$ boundaries are classified as 
    `spheroids' (red). Those with $n_g$ less than 2.0 and placement outside the 
    1$\sigma$ boundaries are classified as `discs' (blue). The rest are kept 
    separately, shown in green colour. {\bf Distribution of structural types} of 
    galaxies is shown with respect to the slope of their dust attenuation curves, 
    where the slopes are either normalised ($A_{FUV}/A_V$, central plot) or 
    reddening-normalised ($R_{FUV}=A_{FUV}/(A_B-A_V)$, right-most plot). The 
    black error-bar in both plots marks the error associated with the x-axis 
    parameter. Discs support lower values of the slope, i.e., exhibit shallower 
    curves than spheroids.}
    \label{fig:classify}
\end{figure*}

\begin{figure*}
	\includegraphics[width=58 mm]{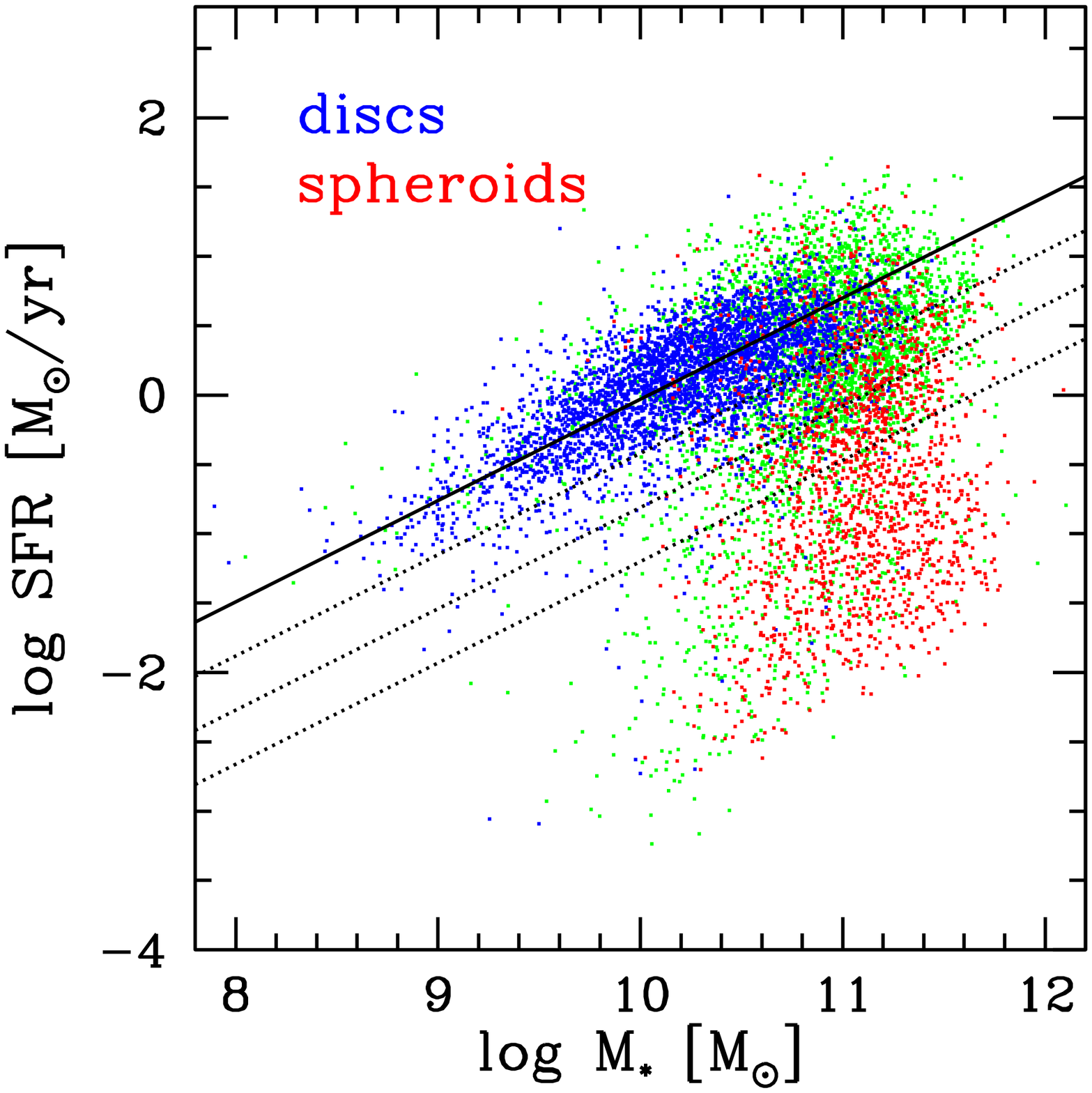}
	\includegraphics[width=58 mm]{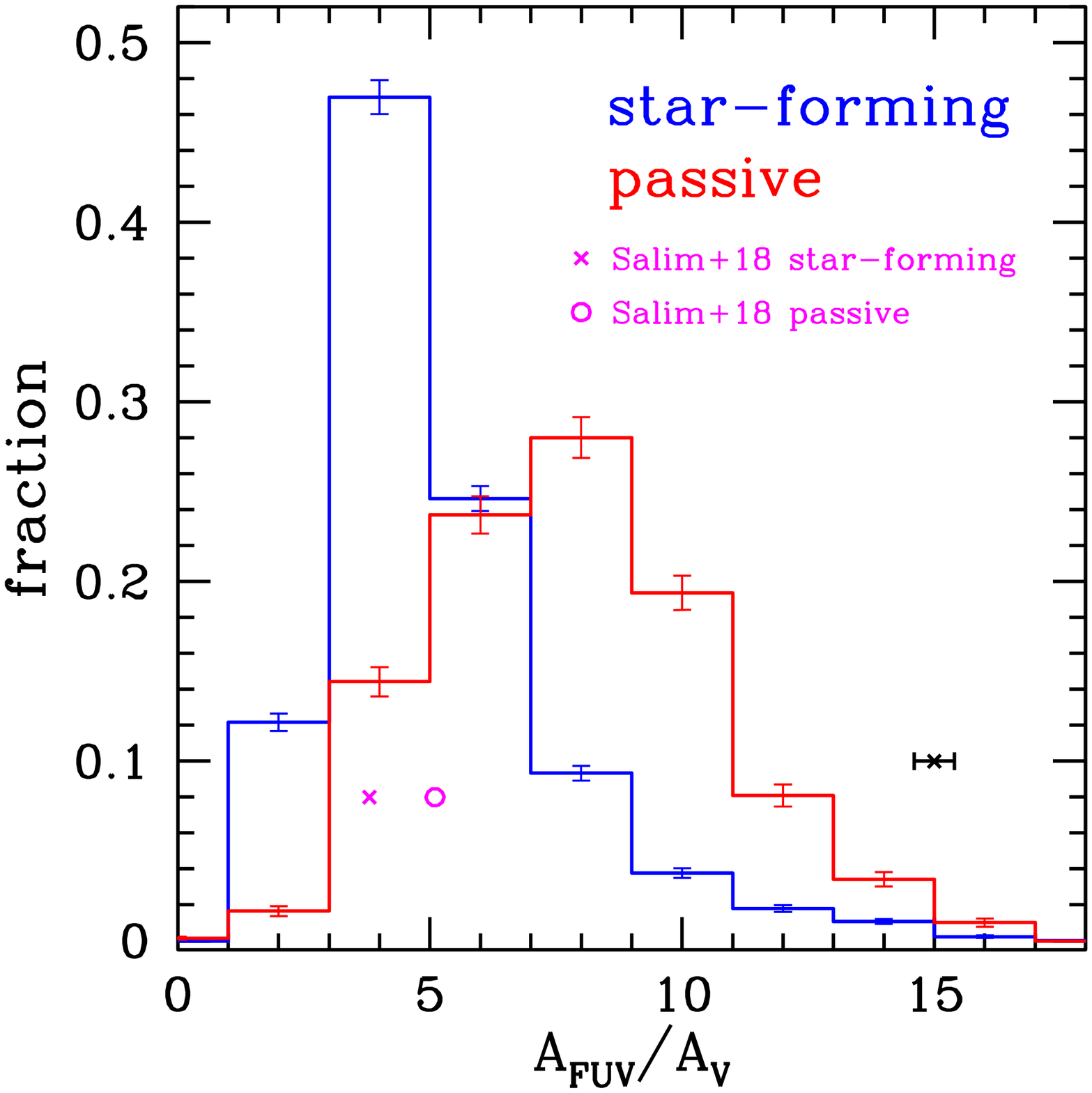}
	\includegraphics[width=58 mm]{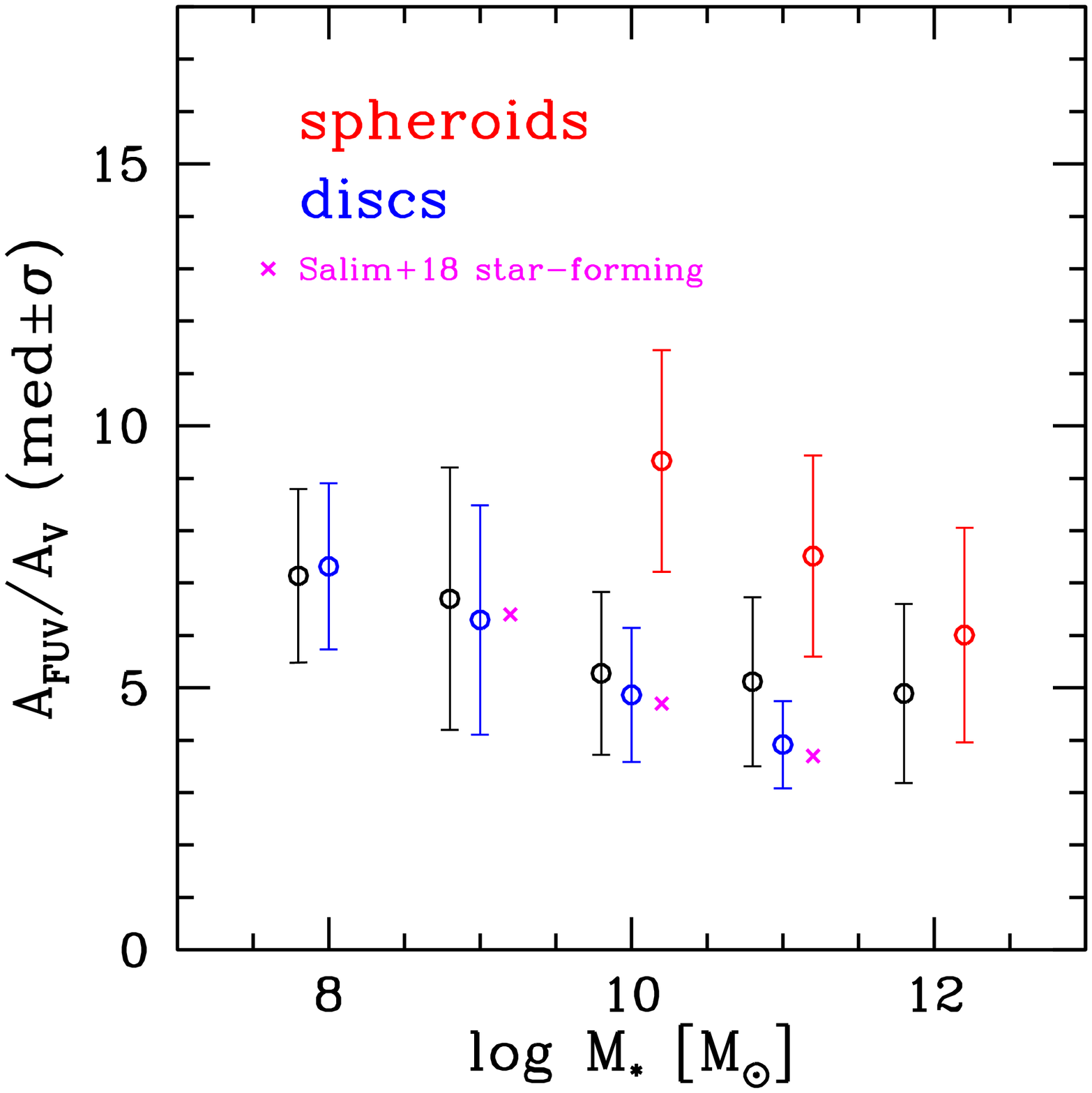}
    \caption{{\bf Distribution on the Main Sequence} (MS) of different structural 
    types, i.e., discs (blue), spheroids (red) and the rest (green), is shown in the 
    left-most plot. The MS relation for local galaxies \citep{Belfioreetal18} and 
    its lower $\sigma$ boundaries are marked with solid and dashed lines. The 
    placement signifies that most ($>85\%$) discs can be marked as star-forming and 
    spheroids can be marked as passive. {\bf Distribution of star-forming and 
    passive galaxies} in our sample is shown with respect to the normalised slope 
    of their attenuation curves ($A_{FUV}/A_V$) in the central plot. It is similar 
    to the distribution observed in the case of discs and spheroids. The average 
    slope values for the two types according to \citet{Salimetal18} have also been 
    marked. The black error-bar marks the error associated with the x-axis parameter. 
    {\bf Median values of the normalised slope} ($A_{FUV}$/$A_V$) are shown in 
    different stellar mass ranges in the right-most plot for our full sample (black), 
    discs (blue) and spheroids (red). The error-bars mark the median absolute 
    deviation associated with each value. The average values from \citet{Salimetal18} 
    for star-forming galaxies (magenta) match those obtained for our disc sample.}
    \label{fig:compare}
\end{figure*}

\begin{figure*}
	\includegraphics[width=58 mm]{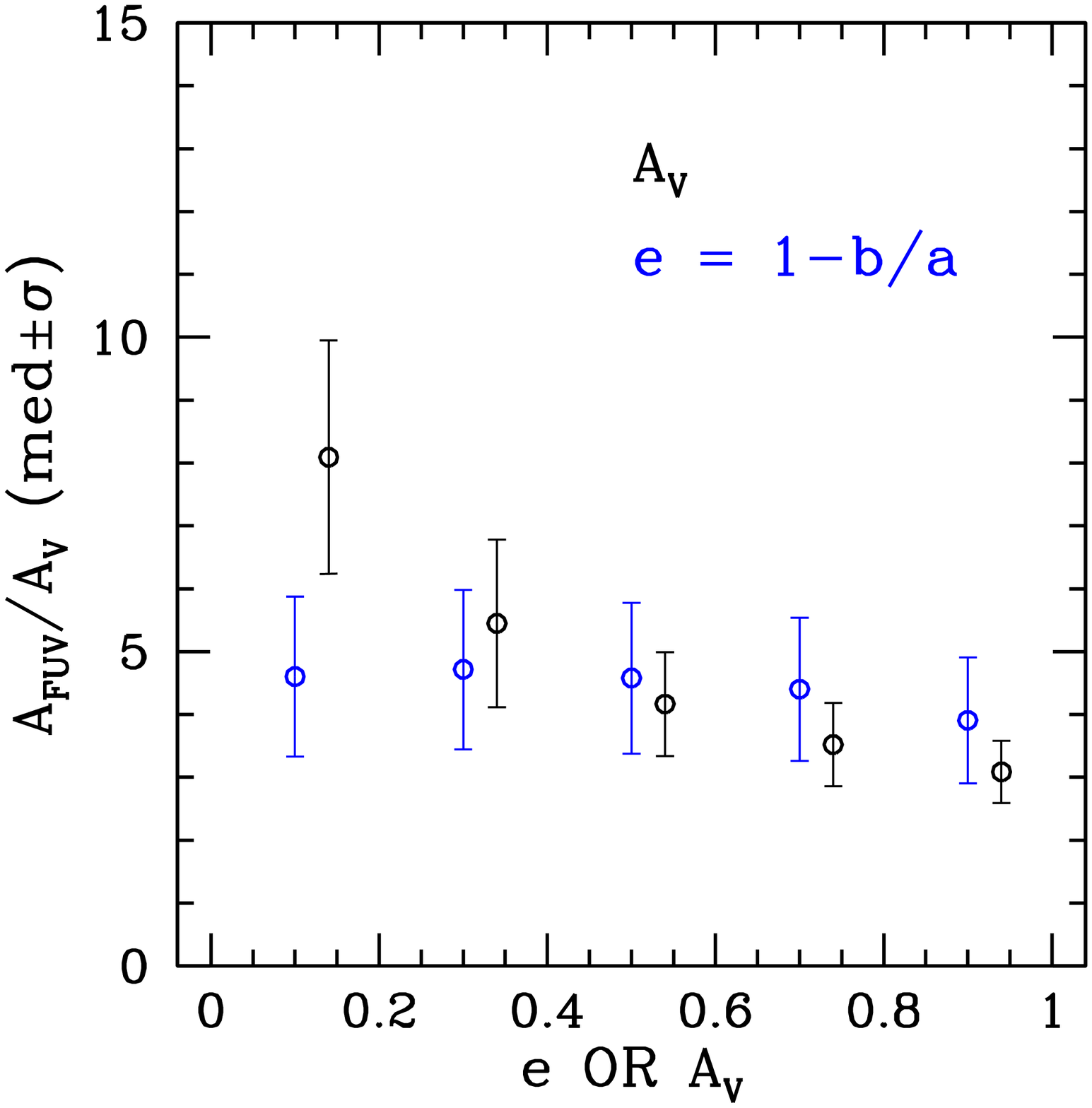}
	\includegraphics[width=58 mm]{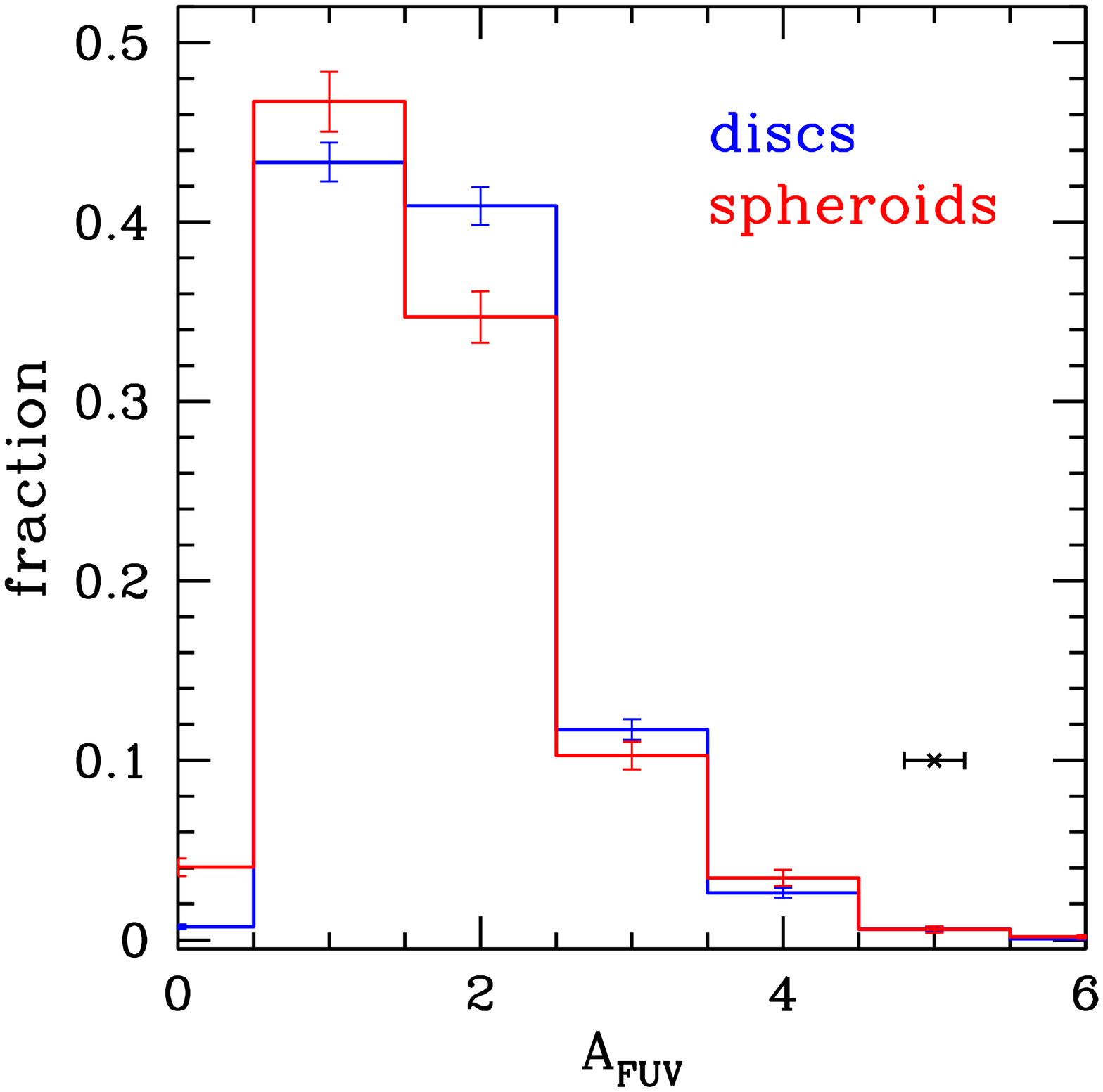}
	\includegraphics[width=58 mm]{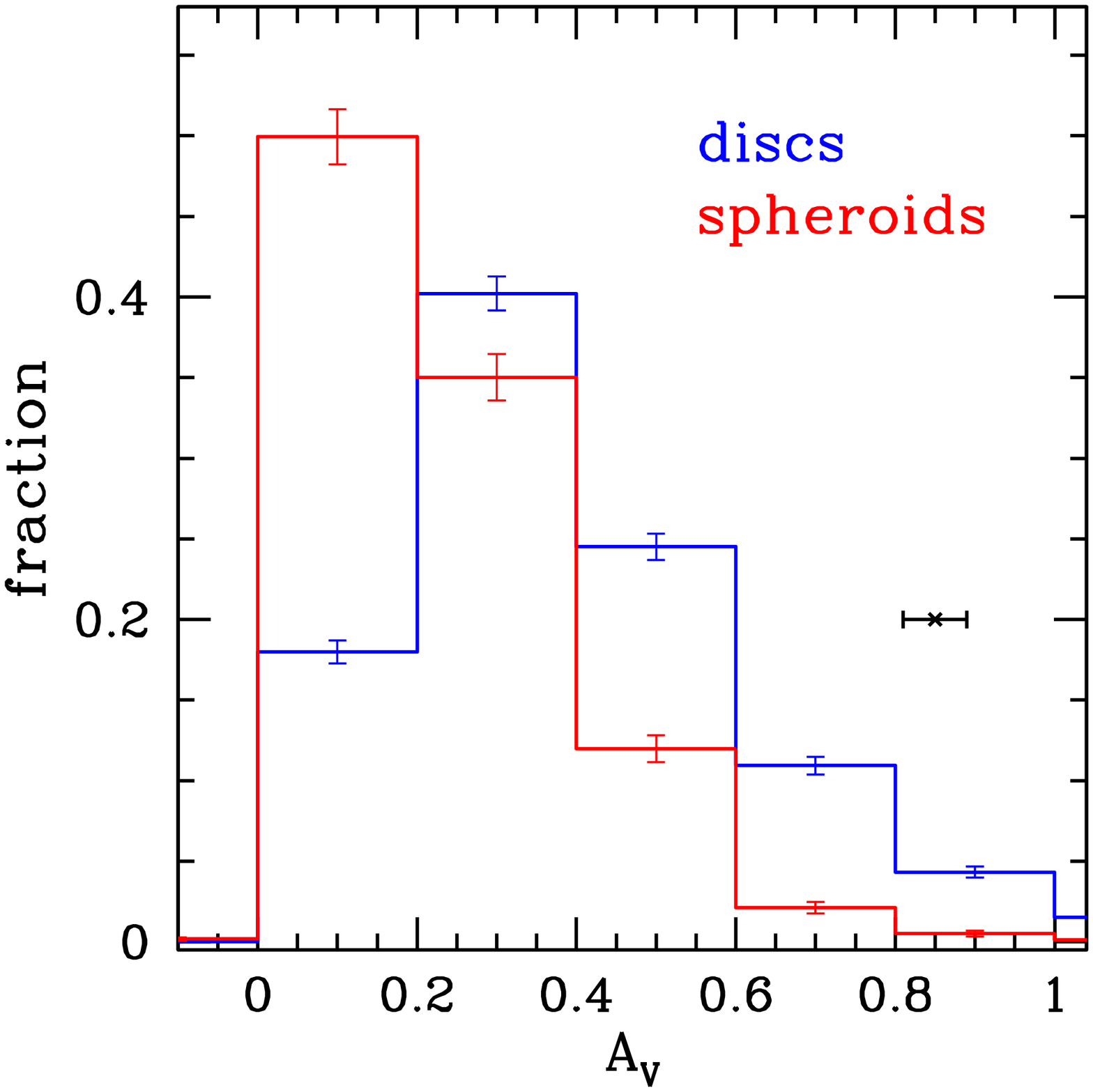}
    \caption{{\bf Median values of the normalised slope} of the attenuation curve 
    ($A_{FUV}$/$A_V$) are shown in the left-most plot for different ranges of the 
    optical attenuation amplitude ($A_V$, in black) and ellipticity ($e=1-b/a$, in 
    blue) of galaxies. The error-bars mark the median absolute deviation associated 
    with each value. While the curve becomes shallower with increasing $A_V$, 
    ellipticity has no visible effect on the slope. {\bf Distribution of attenuation 
    amplitudes} in FUV ($A_{FUV}$, central plot) and optical ($A_V$, right-most plot) 
    are shown for discs (blue) and spheroids (red). The black error-bar marks the error associated with the x-axis parameter in both plots. Discs supporting higher values 
    of $A_V$ compared to spheroids drives the decrease in median values of the 
    attenuation slope (for the full sample) with the increase in $A_V$, as seen in the 
    left-most plot.} 
    \label{fig:dustparam}
\end{figure*}

\section{Results}

To decipher the primary regulators of dust cycle inside a galaxy, we examine the 
dependence of attenuation curve on galaxy's physical properties including structure, 
inclination, total stellar mass and star formation activity. Towards that goal, our 
foremost step is the selection of a representative dust attenuation curve parameter.

\subsection{Dust attenuation curve parameter}

Our primary step is the identification of a dust attenuation curve parameter that is 
representative of its characteristics. The slope of the curve at short wavelengths, 
mainly ultraviolet, is known to be the best in that regard \citep{SalimandNarayanan20}. 
This is driven by the fact that UV is most strongly attenuated by dust. Thus, a 
comparison of attenuation in UV with that in other wavebands is an ideal determinant 
of the effect of interstellar dust on galaxy's stellar emission. Through GSWLC2, we 
have dust attenuation amplitudes in three bands, i.e., FUV, B and V. We, thus, select 
normalised slope of the curve in FUV ($A_{FUV}$/$A_V$) as the defining dust 
attenuation parameter. We find that the trends, reported in this work, become 
stronger with the selection of reddening-normalised slope 
($R_{FUV}=A_{FUV}/(A_B-A_V)$). Nevertheless, to facilitate comparison with other 
studies on the matter, we focus on the normalised slope.

\subsection{Structural classification and correlation}

Prior to the examination of correlation with dust attenuation, we classify the 
structure into two categories, `spheroid' and `disc'. Here, `spheroid' means that the 
galaxy has properties attributed to elliptical galaxies and `disc' means that the 
galaxy has properties attributed to disc galaxies. To classify, we apply two criteria 
concurrently, this includes the placement of galaxies on a projection of the 
fundamental plane, i.e., the Kormendy plane \citep{Kormendy77} and the value of their 
global S\'ersic index ($n_g$). Note that both these criteria, in past studies, have 
been found to be complimentary and efficient indicators of a galaxy's dominant 
structure \citep{Gadotti09,Sachdevaetal15,Sachdevaetal17}. In Figure 
\ref{fig:classify} (left-most plot), the placement of our full sample (8707 galaxies) 
is shown on the Kormendy plane, i.e., global effective radius ($R_{eg}$) versus 
average surface brightness inside that radius ($<$$\mu_{eg}$$>$). Galaxies are 
coloured according to their classification, i.e., spheroids (red), discs (blue) and 
the rest (green).

Based on the classification, we examine the fractional distribution of the structural 
types in different bins of the normalised slope of the dust attenuation curve 
($A_{FUV}$/$A_V$). As shown in Figure \ref{fig:classify} (central plot), 
the two structural types exhibit distinct distribution of slope values. Most discs 
($>90\%$) have slope value less than 6.5, whereas spheroids ($>70\%$) have slopes 
above this value. Note that the distinction becomes even stronger in the case of the 
reddening normalised slope ($R_{FUV}$), where most discs ($>85\%$) have slope value 
less than 16.0 and spheroids ($>85\%$) have slopes above this value (Figure 
\ref{fig:classify} (right-most plot). For both normalised and reddening-normalised 
slopes, the average value for spheroids is nearly twice that for discs, i.e., their 
curve is twice as steep. The slope values for discs are constrained to a narrow range 
in both cases indicating homogeneous attenuation behaviour within a structural group.

Parameters of a galaxy, determined using optical images, might be biased towards the 
distribution of bright stellar populations. To investigate this possibility, we 
employ a representative subset of the full sample (1263), for which we have 
multi-component structural measurements in the $K_S$ band from our earlier work 
\citep{Sachdevaetal20}. Since $K_S$-band covers near-IR wavelengths, where dust 
attenuation is significantly lower than optical wavelengths, it is known to represent 
the underlying mass distribution of a galaxy in the best manner. We find that the 
distribution of galaxies based on their attenuation parameters remains unaltered 
between the optical and $K_S$ band. Thus, the inferences drawn are applicable to the 
overall distribution of stellar mass within galaxies.

\subsection{Stellar classification and correlation}

We classify the galaxies based on their star formation activity into two categories, 
`star-forming' and `passive'. In Figure \ref{fig:compare} (left-most plot), the 
placement of our full sample (8707 galaxies) is shown on the main sequence (MS). The 
MS relation known for local star forming galaxies \citep{Belfioreetal18}, and its 
lower $\sigma$ boundaries are marked. We classify the galaxies within the 1$\sigma$ 
boundary as `star-forming' and those outside the 3$\sigma$ boundary as `passive'. We 
find that the classifications pertaining to structure and star formation activity 
result in a similar selection (Figure \ref{fig:compare}, left-most plot). 
The disc sample ($>90\%$) is within the 2$\sigma$ boundaries of the MS relation and 
the spheroid sample ($>85\%$) is outside those boundaries.

It is thus expected that the distinction in the attenuation curve slopes observed in 
the case of structure, will be reflected in the case of star formation activity as 
well. This is indeed seen to be the case in Figure \ref{fig:compare} (central plot) 
where the fractional distribution of star-forming and passive galaxies is shown in 
different bins of normalised attenuation slope ($A_{FUV}$/$A_V$). The attenuation 
curves for passive galaxies are twice as steep as those for star-forming galaxies. 

We have compared our results with those obtained by \citet{Salimetal18} (Figure \ref{fig:compare} central plot). While the average value of the slope for their 
star-forming galaxies ($A_{FUV}/A_V\sim$3.8) matches with our computation, the 
value for passive galaxies ($A_{FUV}/A_V\sim$5.1) is lower than our estimation. 
This difference is due to the fact that we have applied a more stringent criteria 
for the selection of passive galaxies and kept ambiguous cases separately.

\subsection{Correlation with mass, inclination and optical depth}

In Figure \ref{fig:compare} (right-most plot), the median and median absolute 
deviation values of the normalised attenuation slope ($A_{FUV}$/$A_V$) are shown 
for different ranges of total stellar mass ($M_*$). Notably, our earlier finding 
that the median slope value for spheroids is nearly double the value for discs, is 
maintained in each stellar mass bin. This confirms that the difference of attenuation 
curves for the two structural types is driven by the distribution of stellar mass 
rather than its total amount. However, within a structural group, the total amount is consequential, such that larger $M_*$ results in a slightly shallower slope, as 
reported in earlier works \citep{Salimetal18}. The median values of the 
slope in different stellar mass bins for discs matches those computed by 
\citet{Salimetal18} for star-forming galaxies, as shown in Figure \ref{fig:compare} 
(right-most plot).

In addition to mass, we study the trend in the median values of the slope with the 
ellipticity of the disc component (Figure \ref{fig:dustparam}, left-most plot). In 
fitting galaxies, the ellipticity of the whole system gets measured with a much less 
error margin than that of its components \citep{Bottrelletal19}. We thus select this 
global parameter for our disc dominated sample to study the affect of inclination. 
Note that ellipticity (or axial-ratio) of a galaxy is only an indicator 
of its inclination because the conversion involves assumptions regarding the 
thickness of the galaxy disc \citep{UnterbornandRyden08}. As shown in Figure 
\ref{fig:dustparam}, the ellipticity depicts no trend with the median slope values, 
adding to the argument that increase in the dust mass encountered by the galaxy may 
not be the primal cause of variation in attenuation curves of galaxies.

To probe that further, we analyse the trend in median slope values with the 
amplitude of attenuation in the optical band ($A_V$) for the full sample (Figure 
\ref{fig:dustparam}, left-most plot). In agreement with earlier works 
\citep{SalimandNarayanan20}, the curve does becomes shallower with the increase in 
the optical depth. We argue that this trend is also driven by the structure of the 
galaxies. As shown in the Figure \ref{fig:dustparam} (central and right-most plot), 
while spheroids witness minimal attenuation in the optical compared to the ultraviolet, 
discs witness similar attenuation in both bands. Thus, discs - supporting shallower 
slopes - populate on the larger side of $A_V$ and spheroids - supporting steeper 
slopes - populate on the smaller side of $A_V$. Discs, being more gas rich and 
star-forming, have larger $A_V$ compared to spheroids.

\section{Discussion and summary}

We have probed the cause of variation in the dust attenuation curves of galaxies by 
studying the trend of the normalised attenuation slope in ultraviolet with the 
structure, star formation activity, total stellar mass, inclination and optical depth 
of galaxies. The main findings are discussed:

\begin{itemize}
    \item The attenuation curve for spheroid dominated galaxies is twice as steep 
    as that for disc dominated ones. Both structural types occupy non-overlapping 
    ranges of the attenuation slope value (10-20\% overlap). Thus, stellar mass 
    distribution within a galaxy is one of the main determinant of its dust 
    attenuation behaviour. More importantly, within a structural group, i.e., that 
    of discs, slope values are constrained to a narrow range, emphasising the primacy 
    of structure in determining dust properties.
    \item The steepness of the attenuation curve for spheroids is driven by minimal 
    attenuation in the optical bands compared to that in ultraviolet. Thus, only 
    newly formed stars get attenuated in the case of spheroids. In contrast, the 
    shallowness of the curve in the case of discs is driven by near uniform 
    attenuation in ultraviolet and optical bands. Thus, all stars face attenuation in 
    the case of discs, both due to a more dominant presence of `birth clouds' and a 
    denser interstellar medium \citep{CharlotandFall00}.
    \item The trends observed in the case of structure are reflected in the case of 
    star formation activity as well. The slope for passive galaxies is twice as steep 
    as that for star-forming ones. This is expected because most ($>$85-90\%) discs 
    are star-forming and spheroids are passive. Since star-forming galaxies are more 
    gas rich, clumpier and have a larger presence of birth-clouds compared to passive 
    galaxies, it provides further evidence that the distribution of stellar mass and 
    dust inside a galaxy is the main cause of variation in attenuation curves 
    \citep{Inoue05}.
    \item The distinction in the median slope value for spheroids and discs is 
    maintained in each stellar mass range, adding to the argument that structure is 
    the primary determinant of attenuation behaviour. Within a structural group, the 
    attenuation slope becomes shallower with increasing stellar mass. This is in 
    agreement with earlier works that have reported that increase in the dust mass 
    encountered by a galaxy, reduces the steepness of its attenuation slope 
    \citep{Salimetal18,SalimandNarayanan20}.
    \item All our findings add up to the argument that if the extinction curve is 
    assumed to be universal, the distribution of stars and dust, i.e., star-dust 
    geometry, is the primal determinant of attenuation properties of a galaxy. 
    Recently, under similar assumptions, \citet{Narayananetal18} have shown through 
    simulations that star-dust geometry drives the variation in attenuation curves. 
    Our results confirm that.
\end{itemize}

\section*{Acknowledgements}

We are thankful to the reviewer for insightful comments that have improved the quality 
of this work.

\section*{Data availability}

The data that has not been explicitly presented in the paper will be made available upon 
request to the first author.

\bibliographystyle{mnras}
\bibliography{main}

\end{document}